# Pre-edge XANES structure of Mn in (Ga,Mn)As from first principles

Nataliya A. GONCHARUK[1]*, Jan KUČERA[1], Ludvík SMRČKA[1]

[1] *Institute of Physics, Academy of Sciences of the Czech Republic, v.v.i., Cukrovarnická 10, 162 53 Prague 6, Czech Republic*
* Corresponding author. Tel.: +420-220-318-459; fax: +420-233-343-184; e-mail: gonchar@fzu.cz

**The X-ray absorption near-edge structure (XANES) at the Mn K-edge in the (Ga,Mn)As magnetic semiconductor was simulated using the full potential linearized augmented plane wave (FLAPW) method including the core-hole effect. The calculations were performed in the supercell scheme for a substitutional and two tetrahedral interstitial Mn sites. The resulting pre-edge absorption structures show sharp distinction between the spectra simulated for the substitutional and interstitial Mn defects, which is determined mainly by the second nearest neighbor ligands. A single feature is obtained for the substitutional Mn impurity, whereas two peaks appear for both types of interstitial defects. An interpretation of the pre-edge features is proposed.**

**(Ga,Mn)As magnetic semiconductor / X-ray absorption near-edge structure / Substitutional / Interstitial / Defects**

## Introduction

The Mn-doped GaAs magnetic semiconductors are interesting and promising magnetic materials for potential applications in spin electronics (spintronics). The main motivation behind the preparation and investigation of these materials is the man-made combination of both semiconducting/semimetallic and ferromagnetic properties in one physical system [1-3].

Various configurations of Mn dopants affect the magnetic properties of the (Ga,Mn)As system in different ways. There are several possibilities for a Mn atom to be incorporated in GaAs, and three of them have comparable energy [4]. The substitutional Mn atoms, $Mn_{Ga}^{sub}$, occupying Ga sites, act as hole-producing acceptors which contribute to ferromagnetism. The Mn atoms in tetrahedral interstitial positions, surrounded by either As or Ga atoms, $Mn_{As}^{int}$ and $Mn_{Ga}^{int}$, are electron-producing double donors which destroy ferromagnetic states, as they partly compensate the Mn acceptors in the substitutional positions and reduce the number of holes that mediate ferromagnetism [5,6].

The X-ray absorption spectroscopy is widely used in the study of structural, electronic and magnetic properties of crystals [7,8]. This method allows to analyze the local atomic surrounding, position and electronic state of the chosen chemical element in a complex compound. Recently, a few works have been published reporting on computations of XANES at the Mn K-edge in ternary (Ga,Mn)As alloys [9-13], and in related diluted magnetic semiconductors [14-17]. In the present paper we propose an interpretation of XANES spectra of substitutional and interstitial Mn impurities in (Ga,Mn)As based on first-principles simulations by the FLAPW method [18]. In our calculations an isolated defect is represented by a supercell, which should be as large as possible to describe correctly the absorption process. Therefore, we simulate the spectra from different types of defects by separate calculations, considering always only one substitutional or interstitial atom in a supercell.

## Computational model

The theoretical electron band structure of the (Ga,Mn)As magnetic semiconductor and Mn K-edge XANES spectra were simulated using the FLAPW method as implemented in the WIEN2k package [18]. A generalized gradient approximation [19] was employed for the exchange-correlation functional. Brillouin zone integrations were performed using a tetragonal *k*-point mesh of Monkhorst-Pack type [20].

The spin-polarized calculations were performed within a 64-atom supercell based on the zinc-blende GaAs cubic cell with the experimental lattice spacing $a = 5.65$ Å. We always assumed a single Mn impurity inside each supercell. Thus, the supercell with the embedded substitutional/interstitial Mn atom is doped with 3.125/3.03 atomic percent of Mn impurities, respectively. We used unrelaxed positions of the nearest neighbors around the Mn defect since the effect of relaxation on the band structure is weak. In the substitutional position the Mn atom is surrounded by 4 As with the unrelaxed Mn-As bond distance 2.445 Å, the second neighbors are 12 Ga at the distance 3.994 Å. When Mn is in one of the interstitial positions, its first neighbors are either 4 As or 4 Ga, and the second neighbors are either 6 Ga or 6 As, respectively, with bond lengths equal to 2.445 Å and 2.822 Å. Therefore, we expect that the energy difference between the two configurations, $Mn_{Ga}^{sub}$ and $Mn_{As}^{int}$, in which Mn is surrounded by 4 As atoms, is determined mainly by the second nearest neighbors.

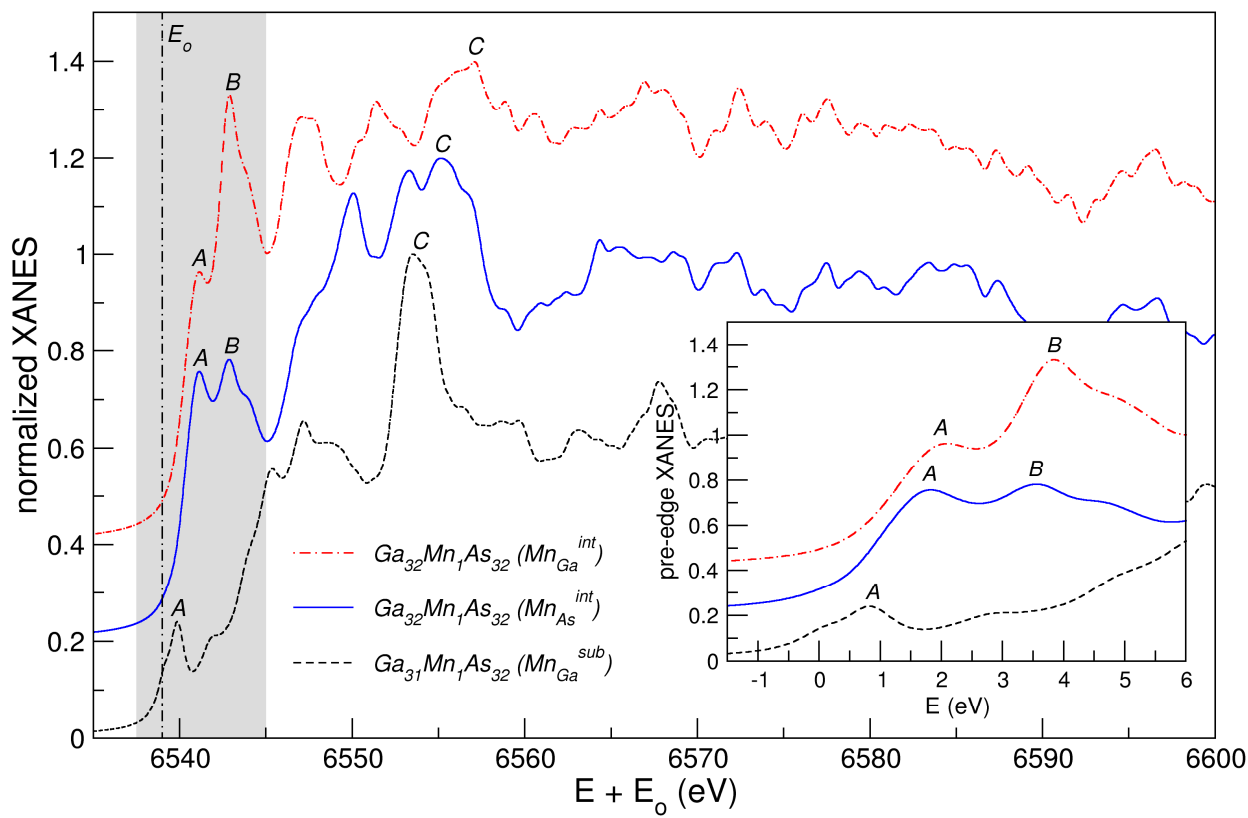

**Fig. 1** Normalized Mn K-edge XANES spectra simulated in the 64-atom (Ga,Mn)As supercell for three different Mn sites in the substitutional ($Mn_{Ga}^{sub}$) and two tetrahedral interstition positions ($Mn_{As}^{int}$ and $Mn_{Ga}^{int}$). Presented spectra are calculated with taking into account the core-hole effect. All spectra are broadened by a Lorentz function with the FWHM $\gamma = 1$ eV. Energies of spectra are shifted to the experimental transition energy of a Mn atom, $E_o = 6539$ eV. The grey region indicates the pre-edge part of absorption spectra, which is shown in detail in the insert. Peaks in the pre-edge structure labeled *A* and *B*, and in the main absorption part labeled *C*, are discussed in the text.

In the single-electron approximation, which is used in our model, the absorption spectrum is given by a transition between the initial ground state with fully occupied core-levels and the final state with one core electron removed and placed in the valence band. The energy of the absorbed radiation is determined as the difference between the calculated total energies of the initial and final states. The core-hole effect is known to be important for reproducing experimental spectra by theoretical calculations [21,22]. Therefore, we carried out our self-consistent band structure calculations for excited states with one core electron removed from the Mn 1*s* level in order to form a core hole after the electron excitation, and an additional valence electron placed at the bottom of the conduction band to fill the lowest unoccupied 4*p* orbital.

Due to the very small radius of Mn 1*s* K-shell core hole the dipole approximation can be safely employed. Then the absorption spectrum is given as a product of the transition probability, calculated from 1*s*-4*p* dipole matrix elements, which is a smooth function increasing with energy, and the local partial density of states (DOS) of 4*p*-states above the Fermi energy in the muffin-tin sphere surrounding the Mn atom, obtained from the final state calculation.

**Results and discussion**

The simulated Mn K-edge XANES spectra for three types of isolated Mn impurity are shown in Fig. 1. All curves were normalized and broadened by a Lorentz function with a full width at half maximum (FWHM) $\gamma = 1$ eV. The use of broadening with larger $\gamma$ would mask the fine details of the absorption pre-edge region, which is the most interesting part of XANES. The spectra were shifted to the experimental transition energy of the Mn atom, $E_o = 6539$ eV.

The XANES spectra were separated into two parts, the pre-edge part (peaks *A* and *B* at $E \sim 6540$ eV) and the main absorption region (the most intense post-edge peak *C* at around 6554 eV), as illustrated in Fig. 1. An intense double structure (peaks *A* and *B*) separated by ~1.8 eV appears in the pre-edge region for both types of interstitial defect, $Mn_{As}^{int}$ and $Mn_{Ga}^{int}$, in contrast to a weak single peak *A* for a substitutional Mn impurity, $Mn_{Ga}^{sub}$.

The arrangement of neighboring atoms around the absorbing Mn atom strongly affects the pre-edge structure. The intensity of pre-edge peaks is much larger for interstitials compared to that for the substitutional model spectrum. In accordance with Wong [23], we attribute this distinction to the different distance to the second neighbor shell ligands coordinating the X-ray absorbing Mn centers. As already mentioned, all considered Mn impurities are located inside tetrahedrons composed of either As or Ga atoms with the same bond length to the first neighbor shell ligands. However, the bond distance of the second nearest neighbor ligands of Mn interstitials is by 1.172 Å less than that of the substitutional Mn. In Wong's terminology, the smaller the "molecular cage", the higher the intensity of the pre-edge absorption. This phenomenon was also observed in our calculations.

The band structure of (Ga,Mn)As with $Mn_{Ga}^{sub}$ calculated with the account for the core-hole effect is shown in Fig. 2. In the supercell scheme with a single defect in a supercell the concentration of defects is finite. The position of the Fermi energy on the energy scale depends on the defect type. In the case of the substitutional Mn impurity it is fixed by the broadened acceptor level near the top of the valence band.

When a Mn atom is in a substitutional position it is surrounded by four As atoms located on the vertices of a tetrahedron. The tetrahedral crystal field of As ions splits 3*d*-states of Mn, which are mainly located in the band gap, into $e_g$- and $t_{2g}$-levels with $e_g$ below $t_{2g}$ [24-26]. Exchange interactions further split these states into spin-up ($\uparrow$) and spin-down ($\downarrow$) states. Due to the tetrahedral arrangement of the ligands the $t_{2g}$-states of the Mn atom hybridize with Mn 4*p*-states. The states resulting from hybridization of the deeper $t_{2g}^\uparrow$-states show mainly Mn *d* character, while those resulting from higher $t_{2g}^\downarrow$-states exhibit dominantly As *p* character. In contrast, in the $e_g$-levels of Mn there are no states available for significant coupling since the GaAs host does not contain $e_g$-states localized in this energy range. It is seen in Fig. 2 that the density of Mn *p*-states is larger at the energy of $d(t_{2g})$-states in comparison to that at the energy of $d(e_g)$-states.

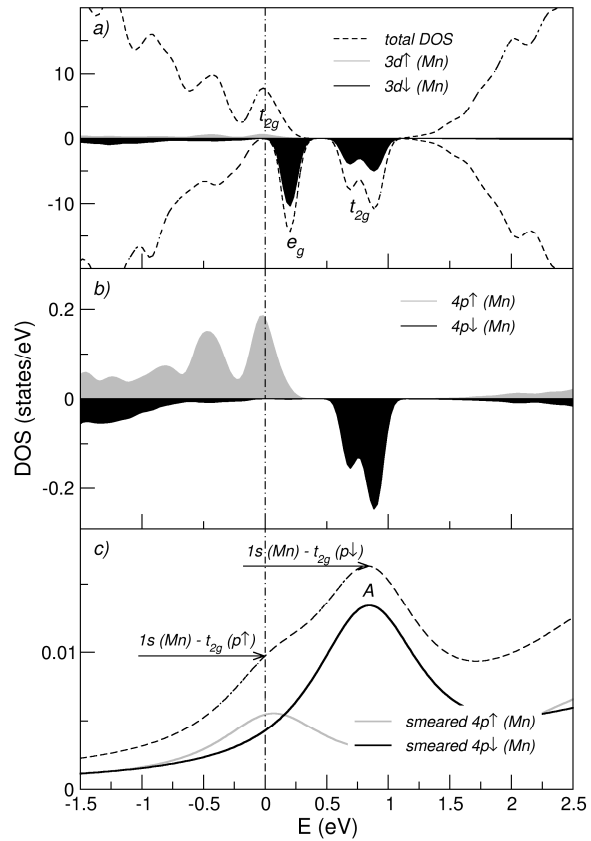

**Fig. 2** The total and partial spin-dependent DOSs calculated with the core-hole effect in the 64-atom supercell with a Mn impurity in the substitutional site. The majority/minority DOSs are presented in the upper/lower part of figures (*a*) and (*b*). The vertical dash-dot line denotes the position of the Fermi level. The total DOSs are shown by dashed curves in (*a*). The density of 3*d*- and 4*p*-states of Mn are illustrated by filled regions in (*a*) and (*b*). The smeared density of 4*p*-states of Mn, i.e., the X-ray absorption spectrum divided by the corresponding matrix element, is presented in (*c*). The electron transitions are depicted by arrows in (*c*).

It is obvious from the discussion above that the pre-edge fine structure of Mn K-edge XANES with $Mn_{Ga}^{sub}$ originates from the transfer of 1*s* electrons to valence 3*d*-states mediated by 1*s*-4*p* dipole transition and 4*p*-3*d* hybridization. We observed merging of two absorption lines corresponding to $1s\text{-}t_{2g}^\uparrow$ and $1s\text{-}t_{2g}^\downarrow$ electron transfers into one line *A* (see Fig. 2(c)). The transfer $1s\text{-}t_{2g}^\downarrow$ is dominant since Mn $t_{2g}^\downarrow$-states above the Fermi level are completely empty. The transfer $1s\text{-}t_{2g}^\uparrow$ becomes impossible as the Fermi level touches the $t_{2g}^\uparrow$-band border, and Mn $t_{2g}^\uparrow$-states are fully filled with electrons.

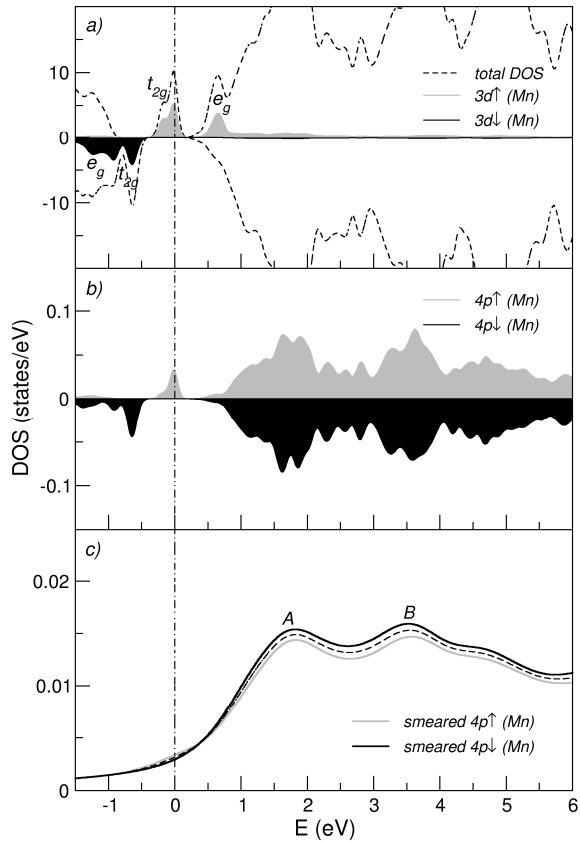

**Fig. 3** The same as in Fig. 2 for the 64-atom supercell with an interstitial Mn impurity inside an As tetrahedron.

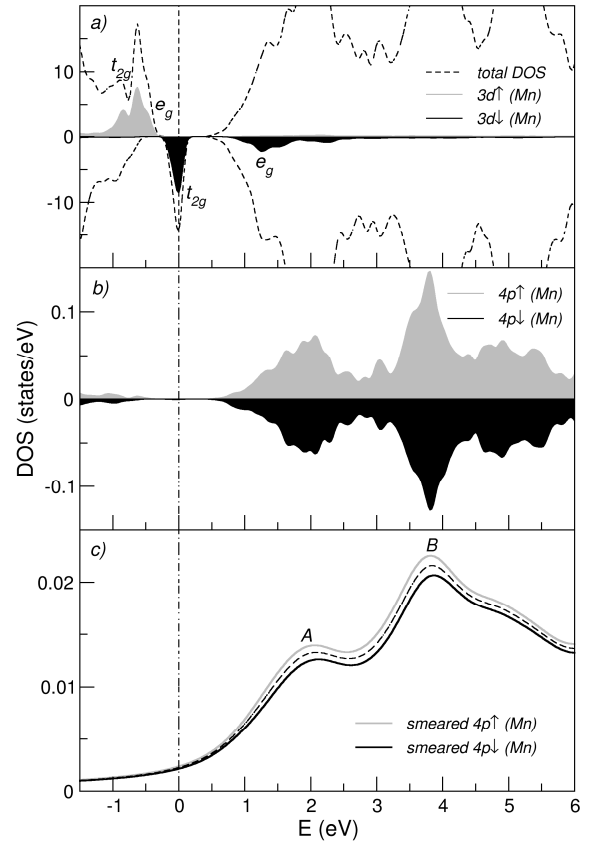

**Fig. 4** The same as in Fig. 2 for the 64-atom supercell with the interstitial Mn impurity inside the Ga tetrahedron.

The band structures of (Ga,Mn)As calculated for either $Mn_{As}^{int}$ or $Mn_{Ga}^{int}$ embedded interstitials with accounting for the influence of the core-hole effect, is shown in Figs. 3, 4. For both interstitials, we observe two intense pre-edge absorption lines *A* and *B*, which are originated from the $1s$-$4p$ transitions. This is in contrast with the pre-edge structure of $Mn_{Ga}^{sub}$. The Fermi energy separates occupied and unoccupied states. In the case of interstitial impurities it is fixed by the broadened donor level close to the bottom of the conduction band. The Fermi level falls in the $t_{2g}^{\uparrow}$-band for $Mn_{As}^{int}$ and $t_{2g}^{\downarrow}$-band for $Mn_{Ga}^{int}$. However, these two bands do not contribute to the pre-edge absorption structure due to low density of Mn $3d$-states near the Fermi level. In contrast to the case with the substitutional Mn, the $t_{2g}^{\uparrow}/t_{2g}^{\downarrow}$-states near the Fermi level do not participate in $4p$-$3d$ hybridization. Thus, both peaks are mainly of dipolar origin, and their shape replicates the DOS of Mn $4p$-states, as seen in Figs. 3, 4.

In conclusion, we have performed an extensive FLAPW numerical study of the K-edge XANES spectra of the substitutional and two tetrahedral interstitial Mn sites in the 64-atom (Ga,Mn)As supercells. We studied weak lines in the pre-edge part of XANES, where the sharp distinction between the spectra with substitutional and interstitial Mn defects was observed. The difference between the simulated spectra is determined mainly by the second nearest neighbor ligands. Two peaks appear in the pre-edge region for any interstitial defect, whereas a smaller single peak is obtained for a substitutional Mn impurity. All pre-edge peaks are small in intensity as compared to the corresponding highest peak C in the XANES main part.

**Acknowledgements**

This work has been supported by the Ministry of Education of the Czech Republic (Center for Fundamental Research LC510), and the Academy of Sciences of the Czech Republic project KAN400100652.

**References**

[1] Y. Ohno, D.K. Young, B. Beschoten, F. Matsukura, H. Ohno, D.D. Awschalom, *Nature* 402 (1999) 790.

[2] H. Ohno, D. Chiba, F. Matsukura, T. Omiya, E. Abe, T. Dietl, Y. Ohno, K. Ohtani, *Nature* 408 (2000) 944.
[3] M. Covington, *Science* 307 (2005) 215.
[4] J. Mašek, F. Máca, *Phys. Rev. B* 69 (2004) 165212.
[5] T. Jungwirth, J. Sinova, J. Mašek, J. Kučera, A.H. MacDonald, *Rev. Mod. Phys.* 78 (2006) 809.
[6] J. Sinova, T. Jungwirth, J. Černe, *Int. J. Mod. Phys.* 18(8) (2004) 1083.
[7] D.C. Koningsberger, R. Prins (Eds.), *X-Ray Absorption: Principles, Applications, Techniques of EXAFS, SEXAFS and XANES*, John Wiley, New York, 1998.
[8] J. Als-Nielsen, D. McMorrow (Eds.), *Elements of Modern X-Ray Physics*, John Wiley, 2001.
[9] R. Bacewicz, A. Twaróg, A. Malinowska, T. Wojtowicz, X. Liu, J.K. Furdyna, *J. Phys. Chem. Solids* 66 (2005) 2004.
[10] A. Titov, X. Biquard, D. Halley, S. Kuroda, E. Bellet-Amalric, H. Mariette, J. Cibert, A.E. Merad, G. Merad, M.B. Kanoun, E. Kulatov, Yu.A. Uspenskii, *Phys. Rev. B* 72 (2005) 115209.
[11] A. Titov, E. Kulatov, Yu.A. Uspenskii, X. Biquard, D. Halley, S. Kuroda, E. Bellet-Amalric, H. Mariette, J. Cibert, *J. Magn. Magn. Mater.* 300 (2006) 144.
[12] F. d'Acapito, G. Smolentsev, F. Boscherini, M. Piccin, G. Bais, S. Rubini, F. Martelli, A. Franciosi, *Phys. Rev. B* 73 (2006) 035314.
[13] A. Wolska, K. Lawniczak-Jablonska, M.T. Klepka, R. Jakieła, J. Sadowski, I.N. Demchenko, E. Holub-Krappe, A. Persson, D. Arvanitis, *Acta Phys. Pol. A* 114 (2008) 357.
[14] X. Biquard, O. Proux, J. Cibert, D. Ferrand, H. Mariette, R. Giraud, B. Barbara, *J. Supercond.* 16 (2003) 127.
[15] S. Sonoda, Y. Yamamoto, T. Sasaki, K. Suga, K. Kindo, H. Hori, *Solid State Commun.* 133 (2005) 177.
[16] O. Sancho-Juan, A. Cantarero, G. Martínez-Criado, N. Garro, A. Cros, M. Salomé, J. Susini, S. Dhar, K. Ploog, *Phys. Status Solidi B* 243 (2006) 1692.
[17] S. Wei, W. Yan, Z. Sun, Q. Liu, W. Zhong, X. Zhang, H. Oyanagi, Z. Wu, *Appl. Phys. Lett.* 89 (2006) 121901.
[18] P. Blaha, K. Schwarz, G.K.H. Madsen, D. Kvasnicka, J. Luitz, *An Augmented Plane Wave plus Local Orbitals program for calculating crystal properties*, In K. Schwarz (Ed.), *WIEN2k*, Techn. Univ. Wien, Austria, 2007, ISBN 3-9501031-1-2.
[19] J.P. Perdew, K. Burke, M. Ernzerhof, *Phys. Rev. Lett.* 77 (1996) 3865.
[20] H.J. Monkhorst, J.D. Pack, *Phys. Rev. B* 75 (2007) 17443.
[21] T. Yamamoto, T. Mizoguchi, I. Tanaka, *Phys. Rev. B* 71 (2005) 245113.
[22] S. Nakashima, K. Fujita, K. Tanaka, K. Hirao, T. Yamamoto, I. Tanaka, *Phys. Rev. B* 75 (2007) 174443.
[23] J. Wong, F.W. Lytle, R.P. Messmer, D.H. Maylotte, *Phys. Rev. B* 30 (1984) 5596.
[24] K. Sato, H. Katayama-Yoshida, *Semicond. Sci. Technol.* 17 (2002) 367.
[25] P. Mahadevan, A. Zunger, *Phys. Rev. B* 68 (2003) 075202.
[26] P. Mahadevan, A. Zunger, *Phys. Rev. B* 69 (2004) 115211.